\documentclass[aps,amsmath,prd,showpacs,nofootinbib,amssymb,preprint]{revtex4}

\usepackage{amssymb,graphics,graphicx,color,epstopdf,subfigure,dcolumn,bm,slashed,tabularx,bm,amssymb}

\def\roughly#1{\mathrel{\raise.3ex\hbox
{$#1$\kern-.75em\lower1ex\hbox{$\sim$}}}}

\begin{document}

\title{LHC Signatures for Cascade Seesaw Mechanism}
\author{Chian-Shu~Chen$^{1,3}$\footnote{chianshu@phys.sinica.edu.tw} and Ya-Juan Zheng$^{2}$\footnote{yjzheng218@gmail.com}}
  \affiliation{$^{1}$Physics Division, National Center for Theoretical Sciences, Hsinchu, Taiwan 300\\
$^{2}$CTS, CASTS and Department of physics, National Taiwan University, Taipei, Taiwan\\
$^{3}$Department of physics, National Tsing Hua University, Hsinchu, Taiwan 300}

\date{\today}
\begin{abstract}
Cascade seesaw mechanism generates neutrino mass at higher dimension (5+4$n$) operators through tree level diagram 
which bring the seesaw scale down to TeV and provide collider signatures within LHC reach. In particular, both Type-II 
scalar and Type-III heavy fermion seesaw signatures exist in such a scenario. Doubly charged scalar decays into diboson is
dominant. We perform a thorough study on the LHC signals 
and the Standard Model background. We draw the conclusion that multilepton final state from interplay of doubly charged scalar and heavy fermion can provide 
distinguishable signatures from conventional seesaw mechanisms. 
\end{abstract}

\pacs{14.60.Pq, 12.60.-i, 14.80.-j, 14.80.Cp}
\maketitle

\section{Introduction}
The effective dimension-five Weinberg operator~\cite{Weinberg:1979sa} violates lepton number by two units and provides 
an explanation for the smallness of neutrino masses successfully if neutrinos are Majorana fermions. Three tree-level 
realizations of this operator correspond to three types of seesaw mechanism through exchange of singlet 
fermion~\cite{Minkowski:1977sc}, triplet scalar~\cite{Magg:1980ut}, and triplet fermion~\cite{Foot:1988aq} 
respectively. These theories are often contemplated in the context of the grand unified theories which live in scales 
much higher than the electroweak theories, therefore, are beyond the reach of the present collider experiments. 
Even though, the phenomenological studies aimed at the TeV scale were investigated in the literature. In particular, 
the like-sign dilepton production at the hadron colliders for Type-I and Type-II seesaw mechanisms were addressed 
in~\cite{Datta:1993nm} and~\cite{Muhlleitner:2003me} respectively. On the other hand, leptons plus jets final states 
from fermion triplet decays in Type-III seesaw were also studied in~\cite{Franceschini:2008pz}. 
It was suggested that the discrimination of these underlying theories is possible by utilizing various multi-lepton 
signals~\cite{delAguila:2008cj}. 

Theoretically, several model building methods provide opportunities to investigate the neutrino mass 
generation directly by lowering the scale to the reach of colliders. For example, one may 
introduce extra particles and/or discrete symmetries to generate neutrino masses radiatively~\cite{Ma:1998dn}. 
Alternatively, the so-called "inverse seesaw mechanism" extends the conventional 
seesaw mass matrix with one additional vector-like singlet, $N = N_{R} + N_{L}$, and it turns out the light 
neutrino masses receive double seesaw suppression factor $\epsilon_{L}m^2_{D}/m^2_{N}$ ($m_{D}$, $m_{N}$ 
and $\epsilon_{L}$ are Dirac mass terms between $\nu_{L}-N_{R}$, $N_{L}-N_{R}$ and new $U(1)$ symmetry 
breaking scale associated to $N_{L}$ respectively)~\cite{Wyler:1982dd}. For a small $\epsilon_{L}$ and a TeV 
scale $m_{N}$, one would give the sub-eV neutrino masses without severely tuning Yukawa couplings~\cite{Ma:2009gu}. 
A general extension of the vector-like singlet $N$ to higher multiplets was also proposed in Ref.~\cite{Law:2013gma}. 

Another method to lower the seesaw scale is that the neutrino mass originates from higher dimension 
operators~\cite{Babu:2001ex}. In particular, there is a class of models called "cascade seesaw mechanism"~\cite{Liao:2010cc} 
and a similar idea can be found in \cite{McDonald:2013kca}. In this case the neutrino masses are generated at tree level from a 
diagram as those of Type-I and Type-III seesaw mechanisms by introducing both new scalar and fermion multiplets, 
carrying quantum numbers $(I_{\Phi},Y_{\Phi}) = (n + \frac{1}{2},1)$ and $(I_{\Sigma}, Y_{\Sigma}) = (n+1, 0)$ for 
$n \geqslant 1$\footnote{We use the convention $Q = I_{3} + \frac{Y}{2}$ in this paper.}. 
In this type of model, the lepton number is violated by the mass insertion of 
the neutral component of the new fermion multiplet and the light neutrino masses are generated via a dimension-($5+4n$) 
operator. The higher dimensionality of neutrino mass operator is due to the development of the vacuum expectation 
value (VEV) of higher isospin scalar field $\Phi^{n+\frac{1}{2}}$ (scalar isospin $n + \frac{1}{2}$ is denoted in this way 
hereafter). In order to generate a naturally small VEV for $\Phi^{n+\frac{1}{2}}$ one needs to engineer the scalar potential 
in a way similar to the Type-II seesaw mechanism. In the context of Type-II 
seesaw mechanism the sign of the scalar triplet quadratic term is positive and the lepton number breaking term (trilinear term 
of scalar triplet and Higgs doublet) will trigger the VEV development of scalar triplet. As a result, a seesaw structure 
between the VEV and the mass of the triplet scalar can be obtained. The development of $\langle \Phi^{n+\frac{1}{2}} \rangle$ 
is realized by generating VEVs of a series of isospin multiplets step by step so that the neutrino mass is suppressed by 
intermediated $\Phi$ and $\Sigma$ with masses located within the reach of LHC.     

In this paper we investigate the LHC collider signatures for such a general scenario of tree level seesaw mechanism with 
a heavy fermion exchange. We find one of the general features of the cascade seesaw mechanism is that both the 
Type-II and Type-III seesaw particles exist. 
This paper is organized as follows: we introduce the generic 
idea of cascade seesaw mechanism in Section-II. In Section-III we briefly 
describe the minimal version of the cascade seesaw mechanism, and then we further 
investigate its collider phenomenology in detail in Section-IV. Then we draw the conclusion in Section-V.

\section{Cascade Seesaw Mechanism}
\begin{figure}
  \centering
  \includegraphics[width=0.45\textwidth]{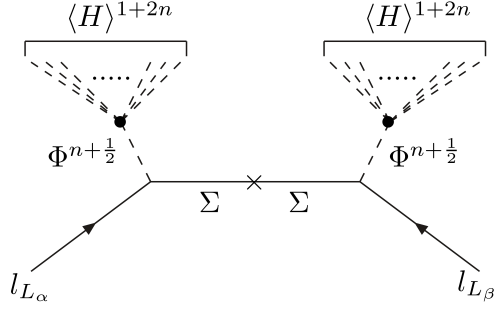}
  \caption{Tree diagram for cascade seesaw mechanism.}\label{dig:numass}
\end{figure}

The requirement of high scale canonical seesaw mechanism is because the neutrino Dirac mass term $m_{D}$ is 
usually considered at the electroweak scale. Naively, if one takes $m_{D}\simeq m_{e}$ the seesaw scale is down 
to ${\cal O}(100)$~GeV, which can be realized by generalizing the tree level seesaw diagram as shown in 
Fig.~\ref{dig:numass}. Yukawa interactions involving the Standard Model (SM) left-handed lepton doublet $l_{L}$ 
are written as 
\begin{eqnarray}\label{yukawa}
{\cal L}_{\rm Yukawa} \supset y_{\alpha\beta}\bar{l}_{L_{\alpha}}Hl_{R_{\beta}} 
+ g_{\alpha i}\overline{l^c}_{L_\alpha}\Phi^{n+\frac{1}{2}}\Sigma_{i} + {\rm h.c.},  
\end{eqnarray}
where $y,g$ are Yukawa couplings, $c$ is the charged conjugation, $H$ denotes the SM Higgs doublet, $\alpha$ 
and $\beta$ refer to flavor indices $e,\mu,\tau$, and $i$ is the new fermion ($\Sigma$) generation index.      
$\Sigma$ field is self-conjugated and can form a Majorana mass term, $\frac{1}{2}\overline{\Sigma^c}M\Sigma + \rm h.c.$, 
expanded as 
\begin{eqnarray}
\overline{(\Sigma^{+(n+1)})^{c}}M\Sigma^{-(n+1)} - \overline{(\Sigma^{+n})^{c}}M\Sigma^{-n} +...+ \overline{(\Sigma^{0})^{c}}M\Sigma^{0} 
+ ...+ \overline{(\Sigma^{-(n+1)})^{c}}M\Sigma^{+(n+1)}.
\end{eqnarray}
We choose a real and diagonal basis here without the loss of generality. The charged Dirac fermions can be defined as 
$\Sigma^{+(n+1)} + (\Sigma^{-(n+1)})^c$, $\Sigma^{+n} - (\Sigma^{-n})^c$, ..., and the neutral Majorana fermion is 
$\Sigma^0 + (\Sigma^0)^c$. In such a way the chiral anomaly is cancelled and the lepton number is violated by the 
Majorana mass term of $\Sigma^0$. The diagram can be divided into two parts, \emph{i.e.} Dirac mass from the second 
term in Eq.~(\ref{yukawa}) and a heavy fermion intermediator. When $\Phi^{n + \frac{1}{2}}$ is taken to be the SM Higgs 
doublet, one can retain the Type-I and Type-III seesaw mechanisms with iso-singlet and iso-triplet fermion as the intermediator 
respectively. 

Different from the canonical seesaws, the black dot vertices in Fig.~\ref{dig:numass} represent how $\Phi^{n+\frac{1}{2}}$ 
develops its VEV via the cascade chain effect involving multi-Higgs external lines.
On the other hand, 
the constraints from electroweak precision measurements put an upper limit on the scalar multiplet VEV. 
Since $\Phi^{n+\frac{1}{2}}$ carries higher isospin, the tree level $\rho$ parameter is given by
\begin{eqnarray}
\rho = \frac{n(n+2)\langle\Phi^{n+\frac{1}{2}}\rangle^2 + \frac{1}{2}(v^2 + \langle\Phi^{n+\frac{1}{2}}\rangle^2)}
{\frac{1}{2}(v^2 + \langle\Phi^{n+\frac{1}{2}}\rangle^2)},
\end{eqnarray}
where we take the SM Higgs VEV $\langle H \rangle = \frac{v}{\sqrt{2}}$. 
From $\rho = 1.0004^{+0.0003}_{-0.0004}$~\cite{pdg}, the first term in the numerator denotes the deviation from 1 
and constrains the $\langle \Phi^{n+\frac{1}{2}}\rangle \lesssim {\cal O}(1)$~GeV. The key point to obtain a small VEV 
of $\Phi^{n+\frac{1}{2}}$ in the cascade seesaw mechanism is the relation between lepton number violation and the 
development of $\langle\Phi^{n+\frac{1}{2}}\rangle$. To make it clear, there exist one renormalizable coupling of scalar 
multiplet to Higgs fields in the potential, which is linear to the lowest isospin state of $\Phi^{n+\frac{1}{2}}$, 
$\Phi^{3/2}\tilde{H}H\tilde{H}$. Here $\Phi^{3/2}$ is the quadruplet by taking 
$n=1$ and this term together with the Yukawa interactions shown in Eq.~(\ref{yukawa}) violates lepton number 
explicitly. The terms relevant to the spontaneous symmetry breaking are 
\begin{eqnarray}
V^{\frac{3}{2}} \supset -\mu^2_HH^{\dagger}H + \lambda_{H}(H^{\dagger}H)^2 
+ \mu^2_{\Phi^{3/2}}\Phi^{3/2\dagger}\Phi^{3/2} - [\kappa\Phi^{3/2}\tilde{H}H\tilde{H} + {\rm h.c.}],  
\end{eqnarray}
where the first two terms are the Higgs potential. The quadratic term of $\Phi^{3/2}$ is set to be positive, 
instead the $\kappa$ term will induce the VEV 
development. As a result, the VEV of $\Phi^{3/2}$ receives a suppression factor, 
$\langle \Phi^{3/2}\rangle \propto v^3/\mu^2_{\Phi^{3/2}}$ 
which is similar to Type-II seesaw mechanism. To generalize this case to higher isospin ($k + \frac{1}{2}$) multiplets, 
one has to utilize the next-to-higher isospin ($k - \frac{1}{2}$) field as a bridge to develop the VEV. For example, the 
quartic term $\Phi^{5/2}\tilde{\Phi}^{3/2}H\tilde{H}$ will induce $\langle \Phi^{5/2}\rangle$ after $H$ and $\Phi^{3/2}$ 
develop VEVs. The procedure can be applied to a sequence of scalar fields $\Phi^{k+\frac{1}{2}}$, 
\begin{eqnarray}
V^{n+\frac{1}{2}} \supset -\mu^2_HH^{\dagger}H + \lambda_{H}(H^{\dagger}H)^2 
+ \sum_{k=1}^{n}\mu^2_{k}\Phi^{k+\frac{1}{2}\dagger}\Phi^{k+\frac{1}{2}} - \sum_{k=1}^{n}[\lambda_{k}(\Phi^{k+\frac{1}{2}}\tilde{\Phi}^{k-\frac{1}{2}}H\tilde{H}) + {\rm h.c.}].
\end{eqnarray}           
Notice that another reason for the positivity of $\mu^2_{k}$ terms is to prevent the tachyonic fields for higher isospin 
multiplets. The general VEV of $\Phi^{n+\frac{1}{2}}$ can be derived as 
\begin{eqnarray}
\langle \Phi^{n+\frac{1}{2}} \rangle = \frac{v^{2n+1}}{2^n\sqrt{2}}\prod_{k=1}^{n}\frac{1}{2\sqrt{2k+1}}\frac{\lambda^*_{k}}{\mu^2_{k}},
\end{eqnarray}   
here the factor $v^{2n+1}$ is corresponding to $2n+1$ Higgs external lines illustrated in 
Fig.~\ref{dig:numass}. It can be read that the VEV of $\Phi^{n+\frac{1}{2}}$ is suppressed by integrating out the 
intermediate scalar multiplets via the cascade chain. Therefore, the general seesaw formula for the neutrino masses 
can be expressed as~\cite{Liao:2010cc}, 
\begin{eqnarray}
m_{\nu_{\alpha\beta}} = \sum_{i}\frac{(-1)^ng_{\alpha i}g_{i\beta}}{(2n+3)M_{\Sigma_{i}}}\frac{v^{4n+2}}{2^{2n+2}}\prod_{k=1}^{n}\frac{1}{2\sqrt{2k+1}}\frac{\lambda^*_{k}}{\mu^2_{k}}.
\end{eqnarray}

\section{The Minimal Model}
It is obvious that the minimal scenario in this class of models appears when $n=1$~ \cite{Kumericki:2012bh}. 
In addition to the SM model particles, we have 
two extra fields, a scalar quadruplet and a fermion quintuplet, 
\begin{eqnarray}
\Phi^{3/2} = (\underbrace{\Phi^{++}, \Phi^{+}, \Phi^{0}}_{{\rm Type-II~seesaw}}, \Phi^{-})^{T} \quad {\rm and} \quad 
\Sigma = (\Sigma^{++}, \underbrace{\Sigma^{+}, \Sigma^{0}, \Sigma^{-}}_{{\rm Type-III~seesaw}}, \Sigma^{--})^{T}.   
\end{eqnarray}
As we can see the scalar triplet in Type-II seesaw and the fermion triplet in Type-III seesaw mechanisms are 
embedded in the particle content. Thus the general features of cascade seesaw mechanism are the coexistence of 
Type-II and Type-III seesaw mechanisms.
The scalar potential is given by 
\begin{eqnarray}
V(H,\Phi^{3/2}) &=& -\mu^2H^{\dagger}H + \lambda(H^{\dagger}H)^2 + \mu^2_{\Phi^{3/2}}\Phi^{3/2\dagger}\Phi^{3/2} 
+ \lambda_{1}(\Phi^{3/2\dagger}\Phi^{3/2})^2 \nonumber \\
&& + \lambda_{2}\tilde{\Phi}^{3/2}\Phi^{3/2}\tilde{\Phi}^{3/2}\Phi^{3/2} + \lambda_{3}H^{\dagger}H\Phi^{3/2\dagger}\Phi^{3/2} 
+ \lambda_{4}\tilde{H}H\tilde{\Phi}^{3/2}\Phi^{3/2} \nonumber \\
&& + (\lambda_{5}\Phi^{3/2}\tilde{H}H\tilde{H} + \lambda_{6}HH\tilde{\Phi}^{3/2}\tilde{\Phi}^{3/2} 
+ \lambda_{7}H\tilde{\Phi}^{3/2}\Phi^{3/2}\tilde{\Phi}^{3/2} + {\rm h.c.}), 
\end{eqnarray}
and it leads the VEV of $\Phi^{3/2}$ to be $\lambda^*_{5}v^3/\sqrt{3}\mu^2_{\Phi^{3/2}}$. The tree level contribution 
to the neutrino mass is obtained,
\begin{eqnarray}
m_{\nu_{\alpha\beta}} = -\frac{1}{6}\lambda^{*2}_{5}\frac{v^6}{\mu^4_{\Phi^{3/2}}}\sum_{i}\frac{g_{\alpha i}g_{i\beta}}{M_{\Sigma_i}},
\end{eqnarray}
with $i$ stands for the number of the fermion quintuplet. For $v = 174$ GeV, $\lambda_{5} = 10^{-3}$, 
$M_{\Sigma}\sim\mu_{\Phi^{3/2}}\sim {\cal O}(10^2)$~GeV and $m_{\nu} \sim 0.1$ eV, we have Yukawa 
couplings around $10^{-2} - 10^{-1}$. The relatively large Yukawa couplings would significantly enhance the search 
probability at the LHC. 
The model would also give flavor changing 
interactions due to the mismatch between the gauge eigenstates and mass eigenstates of neutrinos. The two 
eigenstates can be related by an unitary matrix,
\begin{eqnarray}
\left(\begin{array}{c}\nu_{L} \\\Sigma^0\end{array}\right) = U\left(\begin{array}{c}\nu_{mL} \\\Sigma^0\end{array}\right) \qquad {\rm with} \qquad U = \left(\begin{array}{cc}U_{PMNS} & V_{\nu\Sigma} \\V_{\Sigma\nu} & 1\end{array}\right).
\end{eqnarray}
Here we assume $M_{\Sigma} \gg m_{D}$. The gauge neutral current can be written as 
\begin{eqnarray}
{\cal L}_{NC} &=& \frac{g}{c_{W}}\Big[\frac{1}{4\sqrt{2}}\bar{\nu}\Big(U^{\dagger}_{PMNS}V_{\nu\Sigma}\gamma^{\mu}(1 - \gamma_{5}) 
- V^{T}_{PMNS}V^*_{\nu\Sigma}\gamma^{\mu}(1 + \gamma_{5})\Big)\Sigma^0 \nonumber \\
&& + \frac{\sqrt{3}}{8}\bar{l}V^*_{\nu\Sigma}\gamma^{\mu}(1 + \gamma_{5})\Sigma^{+}\Big]Z_{\mu} + {\rm h.c.},
\end{eqnarray}
and the gauge charged current as
\begin{eqnarray}
{\cal L}_{CC} &=& g\Big[-\frac{\sqrt{3}}{2\sqrt{2}}\bar{\nu}\Big(V^{\dagger}_{PMNS}V_{\nu\Sigma}\gamma^{\mu}(1 - \gamma_{5}) + V^{T}_{PMNS}V^*_{\nu\Sigma}\gamma^{\mu}(1 + \gamma_{5})\Big)\Sigma^{+} \nonumber \\ 
&& - \bar{l}V_{\nu\Sigma}\gamma^{\mu}(1 - \gamma_{5})\Sigma^{0} + \sqrt{\frac{3}{2}}\bar{l}^cV^*_{\nu\Sigma}\gamma^{\mu}(1 + \gamma_{5})\Sigma^{++}\Big]W^-_{\mu} + {\rm h.c.}.
\end{eqnarray}   
Just like the canonical seesaw mechanism, in the minimal realizations the mixing between the light and heavy neutrinos 
is predicted as $V_{\nu\Sigma} \approx \sqrt{m_{\nu}/M_{\Sigma}}$ thus is suppressed. This makes some processes with 
mixing parameters involved difficult to produce at the LHC. However, it is known that a significant mixing between light-heavy 
neutrinos can be obtained as large as $10^{-2}$ which is around the experimental upper limit. Hence the mixings decouple from 
the mass ratio, if $V_{\nu\Sigma}$ is of rank 1 or Tr$(m^T_{D}M^{-1}_{\Sigma^0}m_{D}) = 0$ for three generations of $\Sigma$ 
field~\cite{Buchmuller:1990ds}. In the following section we will study the processes which are insensitive to the mixings.    

\section{Collider Phenomenologies}

\begin{figure}
\begin{centering}
\begin{tabular}{c}
\includegraphics[width=0.5\textwidth]{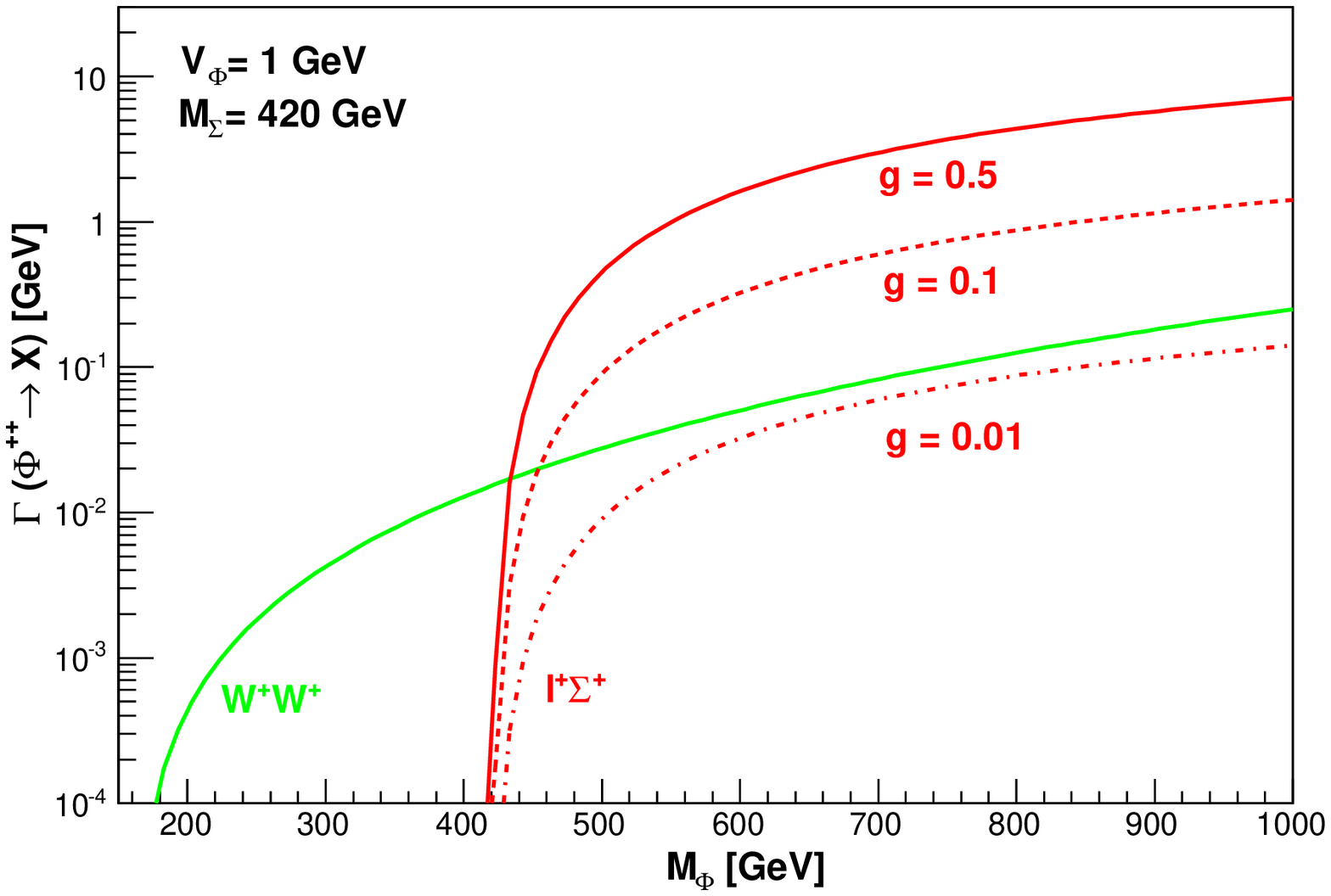}  
\includegraphics[width=0.5\textwidth]{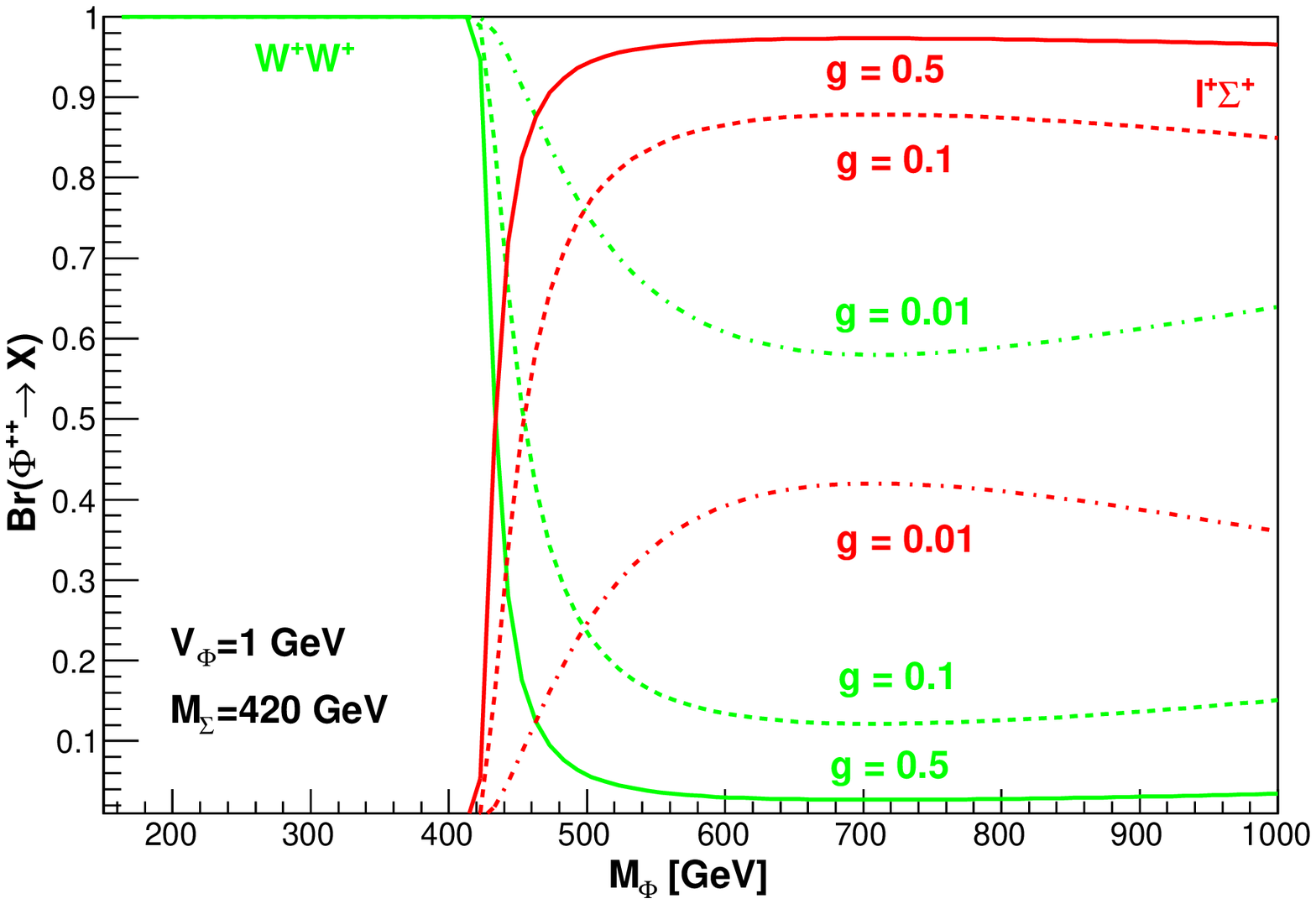}
\end{tabular}
\caption{Decay width and branching ratio of doubly charged scalar decaying into diboson and heavy 
fermion with scalar VEV equals to 1 GeV, heavy fermion to be 420 GeV and Yukawa coupling equals to 0.5, 0.1 
and 0.01, respectively.}
\label{fig:scalardecay} 
\par\end{centering}
\begin{centering}
\par\end{centering}
\centering{} 
\end{figure}

Both Type-II and Type-III seesaw particles exist in the minimal cascade seesaw model.  The doubly charged scalar in 
Type-II seesaw model and the exotic heavy fermion in Type-III seesaw model have been searched at LEP~\cite{LEP}, 
Tevatron~\cite{Tevatron} and the LHC~\cite{LHC,ATLAS:2012hi,Chatrchyan:2012ya}. The up-to-date lower limits 
for the mass of the doubly charged scalar are obtained to be $409$ GeV, $398$ GeV and $375$ GeV if the $100\%$ 
branching ratio of the doubly charged scalar decays into $e^{\pm}e^{\pm}$, $\mu^{\pm}\mu^{\pm}$ and $e^{\pm}\mu^{\pm}$ 
is assumed respectively~\cite{ATLAS:2012hi,Chatrchyan:2012ya}. However, we shall emphasize here that the above 
constraints do not apply to the doubly charged scalar in cascade seesaw models in which the tree level $\Phi^{\pm\pm}$ 
coupling to dilepton is absent. The branching ratios of like-sign dilepton channels are always negligible, which is independent 
of the multi-scalar VEV. The decay channels of $\Phi^{\pm\pm}$ are $W^{\pm(*)}W^{\pm(*)}$ and $\Sigma^{\pm}l^{\pm}$ if 
kinematically allowed. We plot the decay widths and the branching ratios of these two decay channels in Fig.~\ref{fig:scalardecay} 
by taking $M_{\Sigma^{\pm}} = 420$~GeV as the benchmark point~\cite{ATLAS:2013hma}. We find the diboson channel is 
dominant in the case of Yukawa coupling $g=0.01$ and scalar VEV $V_\Phi=1$~GeV. A recent study sets the limits of doubly 
charged scalar mass to be below $43$ GeV and $60$ GeV by using the data at LEP and at the LHC with $7$ TeV collision 
energy and the $4.7$ $\rm{fb}^{-1}$ integrated luminosity. A lower limit is evaluated to be $85$ GeV if the extrapolation of 
the data to 20 ${\rm{fb}^{-1}}$ is used~\cite{Kanemura:2013vxa}. These are the constraints we should adopt in our discussion. 
For the searches of the fermionic triplet lepton in Type-III seesaw model, the CMS collaboration has reported the lower limits 
ranging from $180$ to $210$ GeV in events selected with 3 isolated leptons (the range depends on the selected lepton flavors) 
at $\sqrt{s} = 7$ TeV and an integrated luminosity of 4.9 $\rm{fb}^{-1}$~\cite{CMS:2012ra}. While the ATLAS collaboration 
excludes the heavy lepton mass below 245 GeV in the event selection of at least 4 charged lepton in the final states at 
$\sqrt{s} =8$ TeV and 5.8 $\rm{fb}^{-1}$ of luminosity with the mixing parameter at ${\cal O}(10^{-2})$. Interestingly, the 
probability to have equal to or more than the observed number of events with a background only hypothesis, $p_0$, is 
found to be 0.20 at heavy fermion mass around 420 GeV~\cite{ATLAS:2013hma}. More data accumulation will be helpful 
to demonstrate this signal event. 

\begin{figure}
\begin{centering}
\begin{tabular}{c}
\includegraphics[width=0.5\textwidth]{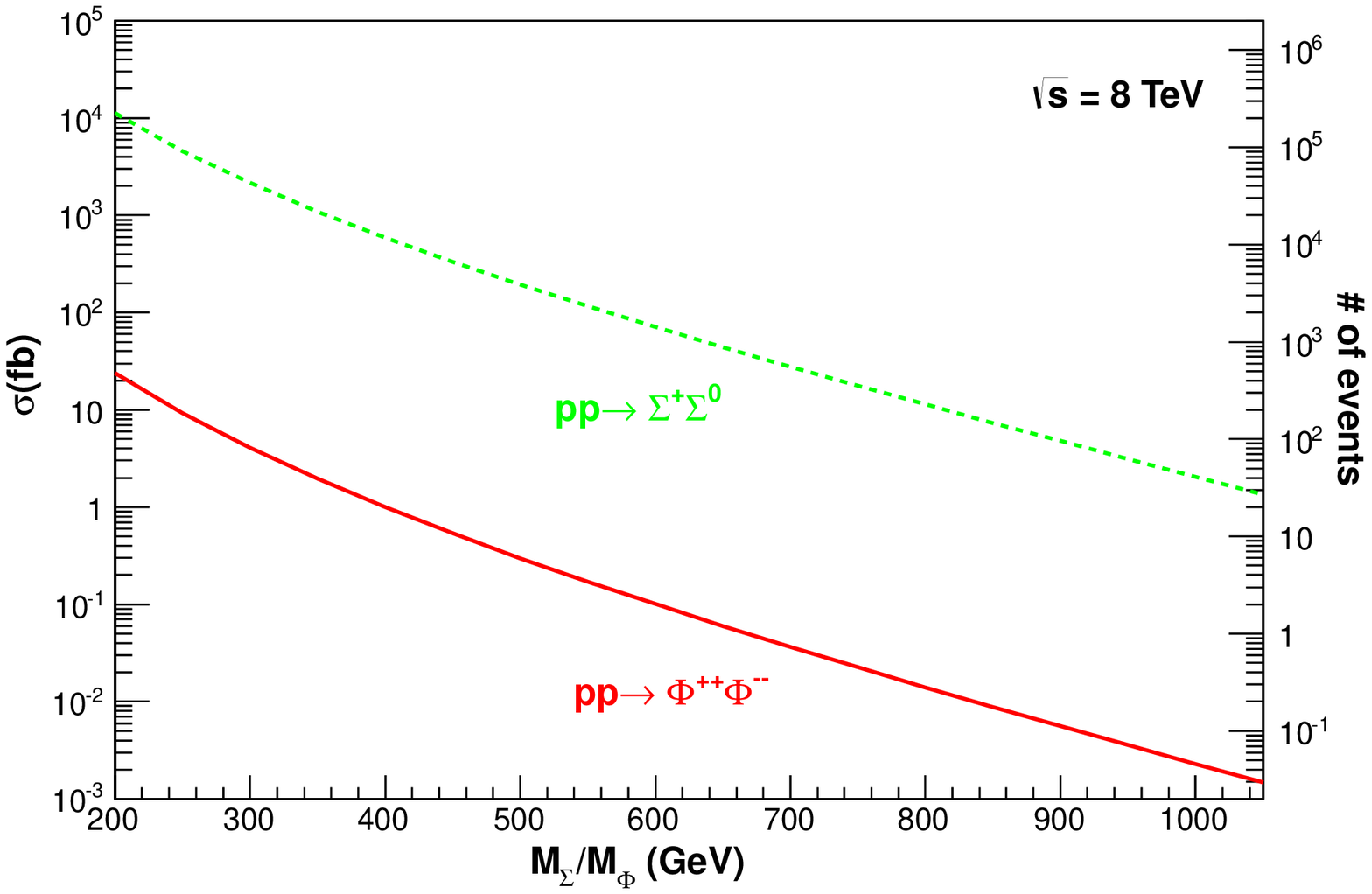}  
\includegraphics[width=0.5\textwidth]{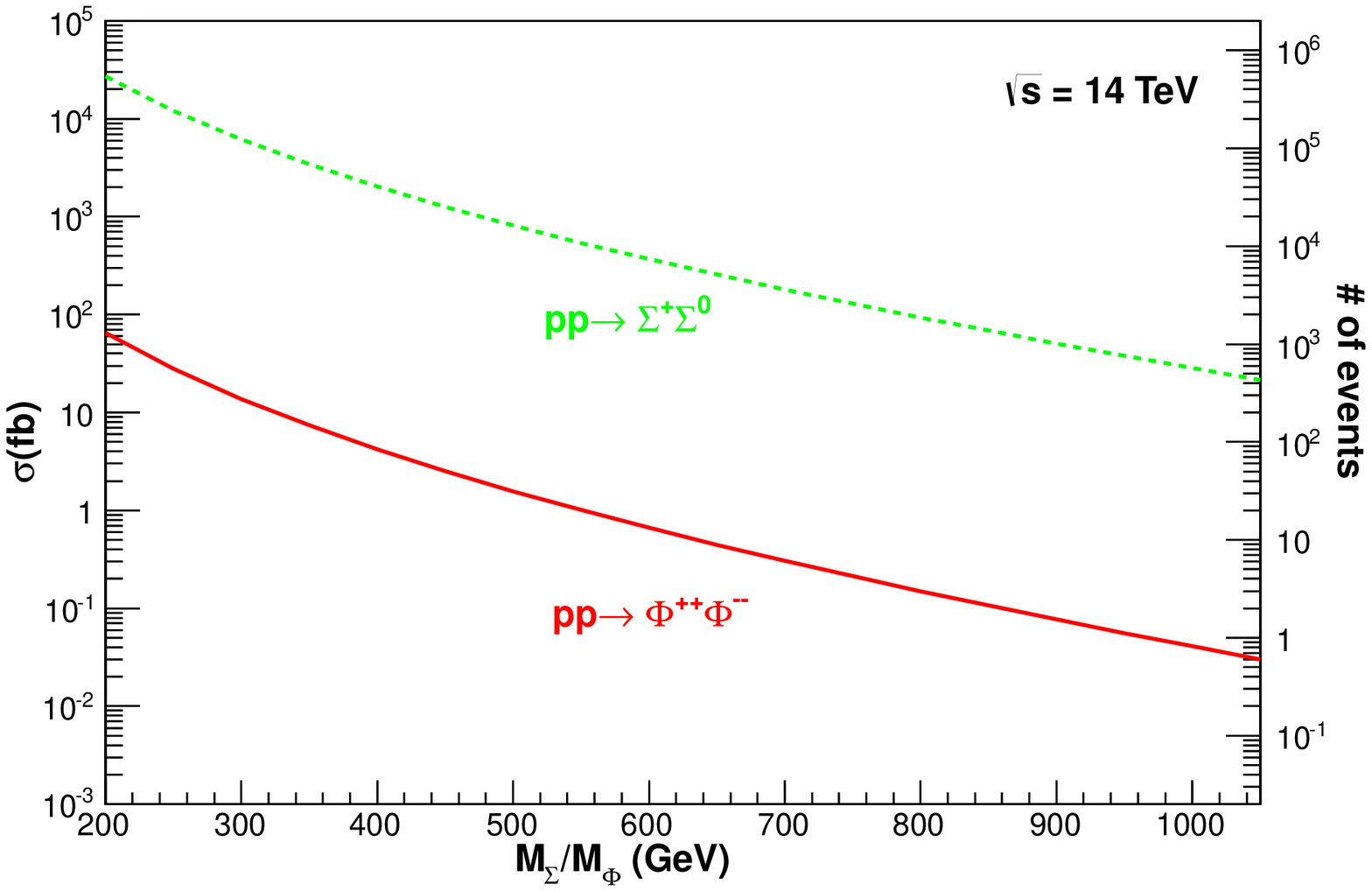}
\end{tabular}
\caption{The cross section for doubly charged scalar and fermion pair production process 
$pp\to \Phi^{++}\Phi^{--}/\Sigma^+\Sigma^0$ with centre of mass energy at 8 (14) TeV luminosity 
assumed to be 20 (300) $\rm{fb}^{-1}$.}
\label{fig:pair} 
\par\end{centering}
\begin{centering}
\par\end{centering}
\centering{} 
\end{figure}

Now we can turn to study the collider signatures of the cascade seesaw mechanism. Throughout the following 
analysis we ignore the mass splittings between the components in the multiplet for simplicity. The pair production 
of both doubly charged scalars $\Phi^{\pm\pm}\Phi^{\mp\mp}$ and exotic heavy leptons $\Sigma^+\Sigma^0$ are 
shown in Fig.~\ref{fig:pair} for 8 TeV (left panel) and 14 TeV (right panel) LHC. Our result is consistent with previous 
studies~\cite{Muhlleitner:2003me,Franceschini:2008pz, delAguila:2008cj,ww}. Comparing the pair production cross 
section between the doubly charged scalars and the heavy fermions, fermion pair production has a larger cross section 
over scalar about ${\cal O} (10^2)$ which is naively because $\Phi^{\pm\pm}\Phi^{\mp\mp}$ is produced via quark-antiquark 
and is much less than quark-quark in the proton-proton collision. At 8 TeV with integrated luminosity assumed to be 20 
fb$^{-1}$, there is still more than 1 number of event expected with the scalar mass up to 600 GeV. For future 
14 TeV run of the LHC, more optimistic number of events can be obtained with 300 fb$^{-1} $ luminosity for both scalar 
and fermion mass up to 1 TeV as shown in the right panel of Fig.~\ref{fig:pair}.  

Since the doubly charged scalar decays into leptons is highly suppressed, we will consider the specific process of 
$pp\to Z^*/\gamma^*\to \Phi^{++}\Phi^{--}\to 2W^+2W^-$ and its signatures at LHC with $M_{\Phi}=300$ GeV. 
The total cross section of this process can be obtained through the general formula
$\sigma=\int f_a(x_{1},Q^2)f_b(x_{2},Q^2)\hat{\sigma}_{q\bar{q}\to 4l+\slashed E_T}(x_{1}x_{2}s)dx_{1}dx_{2}$,
with $\hat{\sigma}_1=\int \frac{1}{2\hat{s}}|{\cal M}|^2 {\rm d}lips_8, 
{\rm and}~ \hat{\sigma}_2=\hat{\sigma}_1(p_1\leftrightarrow p_2)$. $lips_8$ represents the 8-body final state Lorentz 
invariant phase space, $f_a(x_{1})(f_b(x_{2}))$ is the parton distribution function (PDF) of initial state quarks, $\sqrt{s}$ 
is the center of mass energy ($c.m.$) of parton-parton collision, and $\hat{\sigma}$ is the partonic level cross section 
for $q\bar{q}$ process. The eight body final state cross section and contributions from the SM background is also listed 
in Table I. at 8 TeV and 14 TeV LHC.  For the process $\Phi^{++}\Phi^{--}\to 2W^+2W^-\to 4l+\slashed E_T$, 
basic cuts including transverse momentum $p_T$, missing energy $\slashed E_T$, pseudorapidity $|\eta|$ and minimal 
separation $\Delta R_{\rm min}$ cuts are chosen to be  
\begin{eqnarray}
 p_{T}^l>30{\rm GeV},
\slashed E_T>30{\rm GeV},
|\eta_l|<2.5~{\rm and} ~\Delta R_{\rm min}={\rm min}(\sqrt{\Delta \eta^2+\Delta\phi^2})>0.4, 
\end{eqnarray}
respectively according to different distributions of signal and the SM background. To be more realistic, we also perform 
simple detector simulation by smearing the leptons and jets energies according to the assumption
of the Gaussian resolution parametrization 
\begin{eqnarray}
\frac{\delta(E)}{E}=\frac{a}{\sqrt{E}}\oplus b,
\end{eqnarray}
where $\delta(E)/E$ is the energy resolution, $a$ ($b$) is a sampling (constant) term, and $\oplus$ denotes a 
sum in quadrature. We take $a=5\%$, $b=0.55\%$ for leptons and $a=100\%$, $b=5\%$ for jets 
respectively~\cite{Aad:2009wy,Ball:2007zza} and use Madgraph to perform background analysis~\cite{madgraph}. 
We find that in the case of 4-lepton final state at 14 TeV, there is 5 number of events and 1.6 $\sigma$ significance, 
which makes this process observable. With one $W$ decays hadronally, we choose additional basic cuts for 
jets as $p_{T}^j>20{\rm GeV}$, and $|\eta_j|<2.5$. After basic cuts, the significance of $l^+2l^-2j$ process becomes small. 
However, the lower limit on the doubly charged scalar mass is rather weak, and can be as low as 85 GeV in cascade seesaw. 
According to our estimation, the number of events from $l^+2l^-2j+\slashed E_T$ signal can increase 
from ${\cal O}(10)$ to ${\cal O}(10^4)$ for doubly charged scalar mass from 300 GeV to 165 GeV and predict large enough 
significance for observation. It will significantly increase the testability of the model at the LHC. 

\begin{table}[t]
\begin{centering}\scalebox{0.75}{$
\begin{tabular}{|c|c|ccc|cc|}
\hline
[TeV] &{Signal} & \multicolumn{2}{c}{Background}&&$S/B$&$S/\sqrt{S+B}$\tabularnewline
\hline\hline
& ${\bf \Phi^{++}\Phi^{--}\to 2W^+2W^-\to}$ &&&&&\tabularnewline
\hline
&{$  2l^+ 2l^- +\slashed{E_T} $}& $W^+W^+W^-W^-$&$W^+W^-Z$&& & \tabularnewline
 14 &$15.12$&$0.10$ &$30.18$&&$0.50$ &$2.24$  \tabularnewline
   +basic cuts &$5.64$&$-$ &6.73&&$0.838$ &$1.604$\tabularnewline
  \hline
  8 &$0.29$&$0.004$ &$3.4$&&$0.09$&$0.034$ \tabularnewline
\hline
&{$ l^+ 2l^- 2j+\slashed{E_T} $}&$t\bar{t}W^-$& $W^-W^-W^+2j$ & $W^-Z2j$& & \tabularnewline
14 &45 &304.5 & 48.24& 16722&0.003&0.344\tabularnewline
$P_T^{(l,j)}>20(30){\rm GeV}$ & 23&146 &23.8 &7958 &$0.002$&$0.25$\tabularnewline
$\slashed E_T>30{\rm GeV}$ & 22.4&140 &18.8&3392 &$0.006$&$0.37$\tabularnewline
$|\eta_{l,j}|<2.5$ & 19.1&125.3 &13.9&2531 &$0.007$&$0.37$\tabularnewline
$\Delta R>0.4$ & 10.2&122 &11.8&2165 &$0.004$&$0.21$\tabularnewline
\hline
8  & $5.8$& 7.09 & 0.814 & 355.4&0.016&0.302\tabularnewline
\hline\hline
&${\bf \Sigma^+\Sigma^0\to}$&&&&&\tabularnewline
\hline
&$l^+Z^0 l^-W^+\to 2l^+2l^-+2j$&$t\bar{t}Z$&$W^+W^-W^+Z$&$W^+ZZ$&&\tabularnewline
14 &177& 381.9 & 0.1854 & $6.249$&0.46&7.4 \tabularnewline
8 & 4.2& 6.1 & $0.0043$ & $0.1994$&30&2.4\tabularnewline
\hline
 &$l^+Z^0l^+W^-\to 3l^+l^-+2j$&$W^+W^+Z2j$&&&&\tabularnewline
14 &177& $0.4137$&&& 427.8&13.1\tabularnewline
8 &4.2&0.01&&&420&2.04 \tabularnewline
\hline
 & $l^+Z^0 l^-W^+\to 3l^+2l^-+{\slashed E_T}$&$W^+W^-W^+Z$&$W^+ZZ$&&&\tabularnewline
14 & 29.5& $0.0417$ & $2.191$ &&13.2&5.24\tabularnewline
8 &0.7& $0.001$ & $0.076$&&9.1&0.79 \tabularnewline
\hline
\end{tabular}$}\caption{Number of events of scalar $M_{\Phi}=300$ GeV and fermion $M_\Sigma=300$ GeV pair 
production and decay as well as corresponding SM backgrounds at centre-of-mass energy 8 TeV (20$\rm{fb}^{-1}$) and 
14 TeV (300$\rm{fb}^{-1}$). In both signal and backgrounds, charged lepton of $e^-$ and $\mu^-$ are included. In heavy 
fermion decay process, $|V_{l\Sigma}|=g=0.01$ is adopted as mixing parameter and Yukawa coupling.}
\label{tab:cs}
\par\end{centering}\end{table}

For heavy fermions, specifically we choose the same process as the LHC search~\cite{ATLAS:2013hma} and 
study $pp\to W^+(k)\to \Sigma^+\Sigma^0$ decaying into $\Sigma^+\to l^+ Z^0 \to l^+  l^+l^-$ and 
$\Sigma^0\to l^- W^+\to l^-jj^\prime$ i.e., $l^+l^+l^-l^-jj$ final state. Together with other final states as $3l^+l^-2j$ 
which violates lepton number explicitly, $3l^+2l^-\slashed E_T$ which is purely multi-lepton final state, we find that 
the SM background is much smaller than the chosen signal processes, providing clean signatures. The pair production 
of doubly charged fermions is also studied in~\cite{Kumericki:2012bh}. For $2l^+2l^-+\slashed E_T$ final state from the 
doubly charged fermion pair decay, we find that at 8 TeV LHC hundreds of $\Sigma^{++}\Sigma^{--}$ can be produced, which also makes the SM background process $W^{+}W^{+}W^{-}W^{-}$ and $W^{+}W^{-}Z$ negligible as shown in the first process of Table.~\ref{tab:cs} with the same final state. Therefore, heavy fermion production processes provide promising clean channels for LHC study on cascade seesaw mechanism.

\begin{figure}
\begin{centering}
\begin{tabular}{c}
\includegraphics[width=0.45\textwidth]{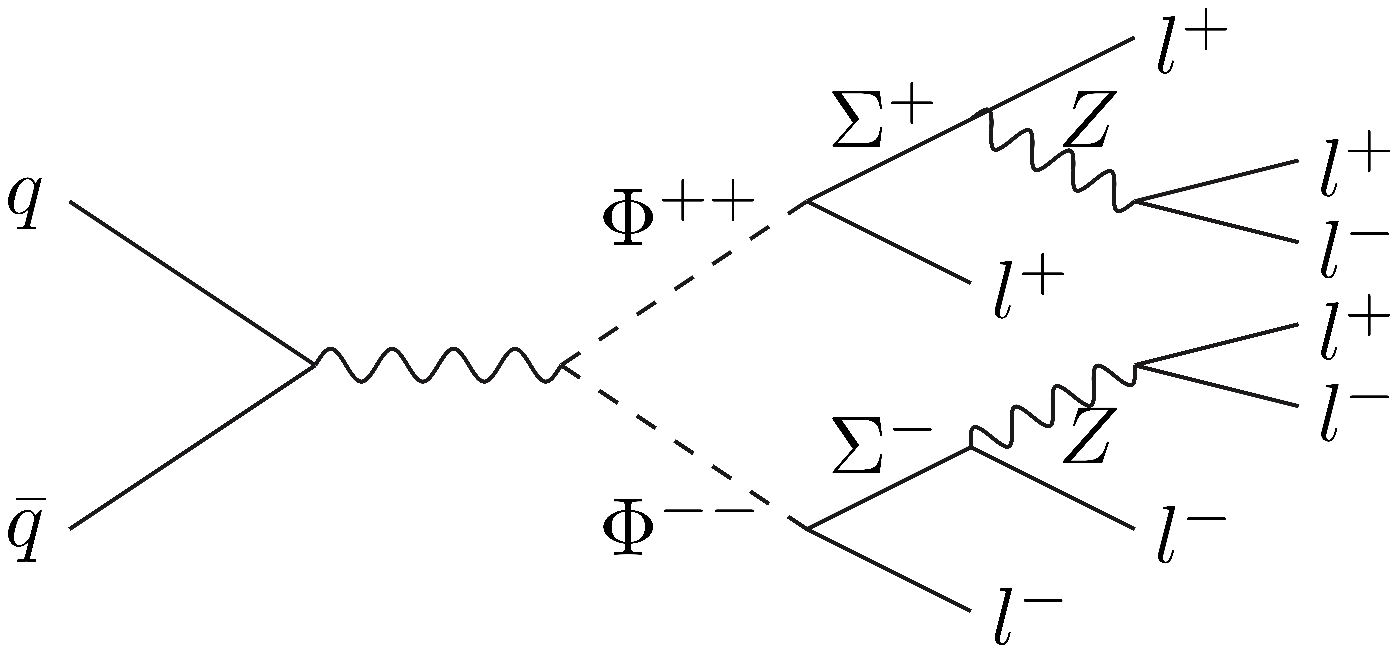}  
\end{tabular}
\caption{The Feynman diagram for $pp\to \Phi^{++}\Phi^{--}\to{\Sigma}^+\Sigma^-l^+l^-\to4l^+4l^-$ with 8-lepton final state.}
\par\end{centering}
\begin{centering}
\par\end{centering}
\centering{} 
\label{8l}
\end{figure}

\begin{figure}
\begin{centering}
\begin{tabular}{c}
\includegraphics[width=0.5\textwidth]{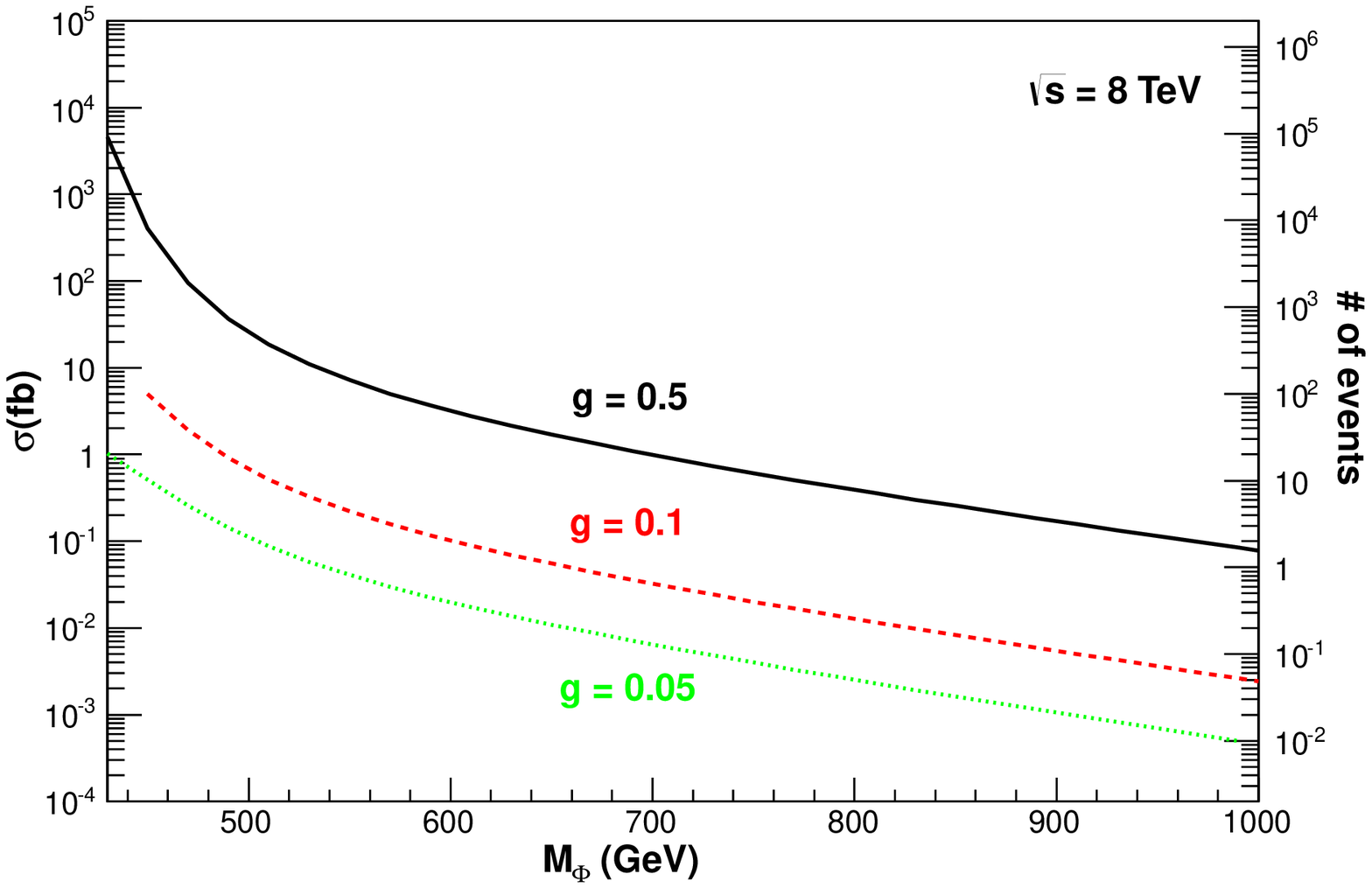}  
\includegraphics[width=0.5\textwidth]{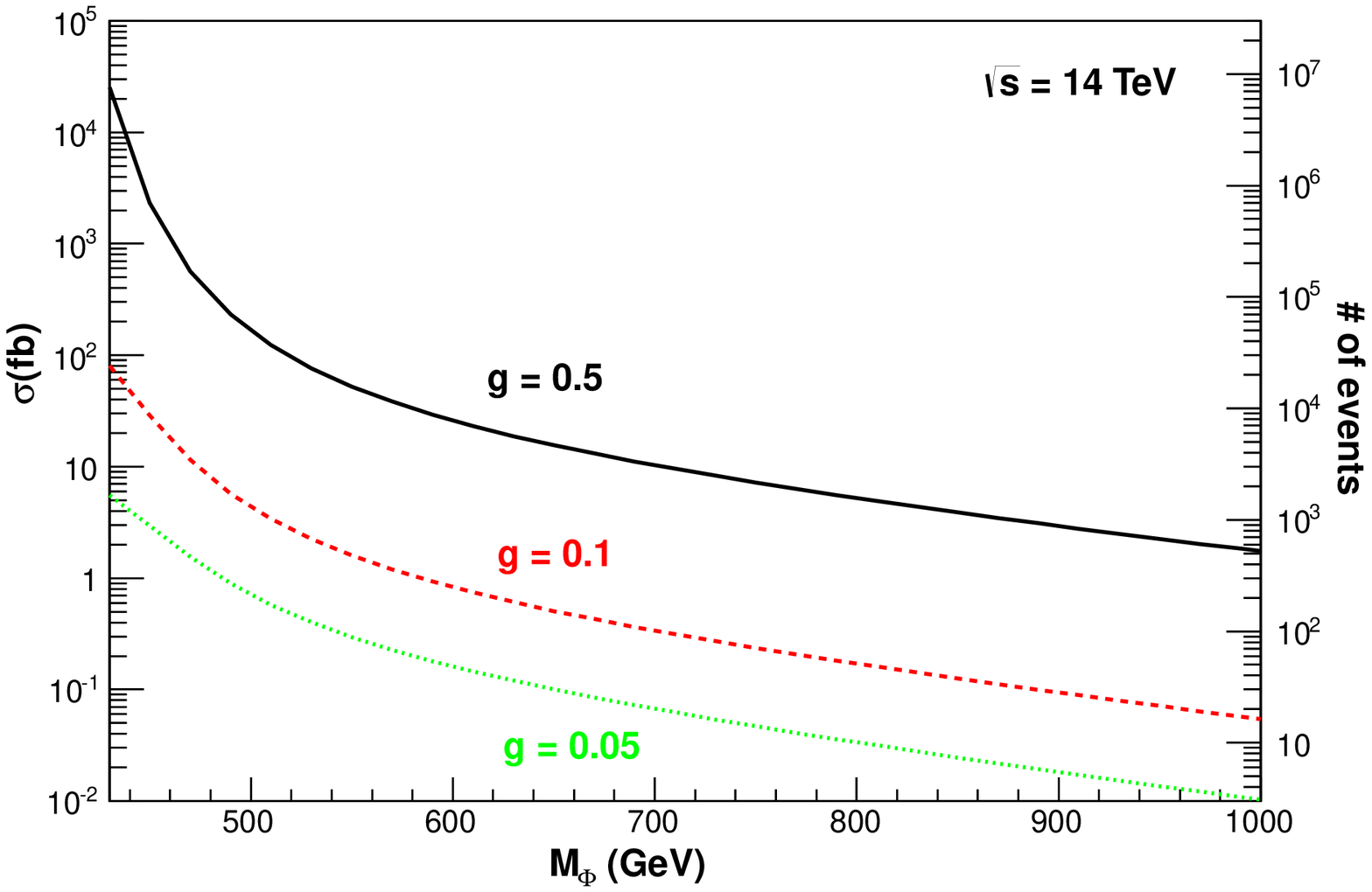}  \tabularnewline
\end{tabular}
\caption{The cross section for $pp\to \Phi^{++}\Phi^{--}\to{\Sigma}^+\Sigma^-l^+l^-\to4l^+4l^-$ at center-of-mass energy equals to 8 TeV (left panel) and 14 TeV (right panel). Different mixing parameters are chosen as Yukawa coupling $g =$ 0.5 (black solid line), 0.1 ( red dashed line) and 0.05 (green dotted line). }\label{fig:cs-ZH} 
\par\end{centering}
\begin{centering}
\par\end{centering}
\centering{} 
\label{4l-cut}
\end{figure}

In addition, larger multiplets in such a model contains more particles which guarantees us distinctive signatures. 
We find a novel process as plotted in Fig. 4 that the doubly charged scalar production and then 
decays into heavy fermions with 8-lepton final state without missing energy. As shown in Fig.~\ref{fig:cs-ZH}, 
$pp\to \Phi^{++}\Phi^{--}\to{\Sigma}^+\Sigma^-l^+l^-\to4l^+4l^-$ cross section as well as number of events with luminosity 20 fb$^{-1}$ for 8 TeV (left panel) and 300 fb$^{-1}$ for 14 TeV (right panel) are displayed. Such a multi-lepton process is very clean from the SM background. The only possible SM background is from $4Z$ decaying leptonically, but according to our estimation it is at order of $10^{-7}$ fb at 14 TeV LHC  thus is negligible.
For illustration, we take $M_\Sigma=420$ GeV Yukawa couplings $g$ =0.05, 0.1 and 0.5 respectively. In this process,  the mixing parameter $|V_{l\Sigma}|^4$ appeared in the coupling can be cancelled by $|V_{l\Sigma}|^4$ in the propagators using narrow width approximation. As a result, this process is insensitive to the mixing parameter between light and heavy fermions.  Obviously, for 8 TeV run of LHC only $g > 0.05$ with scalar mass below 500 GeV we can predict significant signal events for observation. For 14 TeV LHC, with Yukawa coupling approximates to 0.05 and $M_\Phi$ up to 1 TeV, there are still more than 10 number of events expected.

\section{Conclusion}
Higher dimension operators give an explanation to the smallness of the neutrino masses and provide the opportunity to 
probe the origin of neutrino mass mechanism. Cascade seesaw mechanism is based on the spirit of the canonical seesaw 
mechanism with the extension of the scalar and fermion sectors to higher dimension representations. The neutrino masses are 
hence generated at dimension $5+4n$ operators. We review the main consequences of the cascade seesaw mechanism 
in a general form. A novel signature in cascade seesaw models is that both Type-II and Type-III seesaw particles exist. Then we study the LHC signatures for the minimal model with detailed analysis of signals and the corresponding SM background. For the extra scalar multiplets, we discuss the processes with scalar decaying into diboson which is the dominant process for most of the parameter space. Since such a decay channel of doubly charged scalar receives weaker constraints from experiments, the mass can be smaller enough to produce large number of events for observation.  To be consistent with the experimental search on Type III heavy fermion, we study neutral and singly charged heavy fermion production. Clean multi-lepton signatures from SM background are studied. The most non-trivial signature is heavy fermion associated with lepton from doubly charged scalar decay with 8-lepton final state. This process provides a distinctive signal from other seesaw models.

\subsection*{Acknowledgement}
The authors are supported by National Center for Theoretical Sciences, Taiwan, R.O.C. (CC) and NSC of ROC (YZ). They are grateful to Prof. X. He and Prof. Y. Liao for useful discussion on this work. YZ would like to thank KEK for their hostality where part of this work was done.

\end{document}